\documentclass[conference]{IEEEtran}
\usepackage{amssymb}
\usepackage[cmex10]{amsmath}
\usepackage{theorem}
\usepackage{latexsym}
\usepackage{graphicx}
\IEEEoverridecommandlockouts
\theoremheaderfont{\bf}
\theorembodyfont{\sl}
\newtheorem{theo}{Theorem}[section]
{\theorembodyfont{\rm} \newtheorem{defi}[theo]{Definition}}
{\theorembodyfont{\rm} }
{\theorembodyfont{\rm} \newtheorem{exa}[theo]{Example}}
{\theorembodyfont{\rm} }
\newtheorem{prop}[theo]{Proposition}

{\theorembodyfont{\rm} }

{\theorembodyfont{\rm}}
{\theorembodyfont{\rm}}

\newcommand{\F}{{\mathbb F}}
\newcommand{\Z}{{\mathbb Z}}

\newcommand{\CC}{{\mathcal C}}

\newcommand{\SSS}{{\mathcal S}}

\newcommand{\A}{{\mathcal A}}

\newcommand{\cT}{{\mathcal T}}

\newcommand{\cY}{{\mathcal Y}}

\newcommand{\cP}{{\mathcal P}}
\newcommand{\Bf}{{\mathfrak B}}

\newcommand{\zerob}{{\mathbf 0}}
\newcommand{\ab}{{\mathbf a}}
\renewcommand{\sb}{{\mathbf s}}

\newcommand{\inner}[1]{\langle{#1}\rangle}

\newcounter{alp}
\newcounter{ara}
\newcounter{rom}
\newenvironment{romanlist}{\begin{list}{(\roman{rom})\hfill}{\usecounter{rom}
     \topsep0ex \labelwidth.7cm \leftmargin.7cm \labelsep0cm
     \rightmargin0cm \parsep0ex \itemsep.3ex
     \partopsep1.6ex}}{\end{list}}

\begin{document}
%
\title{Reducing complexity of tail-biting trellises}

\author{
\IEEEauthorblockN{Heide Gluesing-Luerssen}
\IEEEauthorblockA{University of Kentucky\\
Department of Mathematics\\
Lexington, KY 40506\\
Email: heide.gl@uky.edu
}
\and
\IEEEauthorblockN{G.~David Forney, Jr.}
\IEEEauthorblockA{Laboratory for Information and Decision Systems\\
Massachusetts Institute of Technology\\
Cambridge, MA 02139\\
Email: forneyd@comcast.net
}

\thanks{The work of the second author was supported in part by National Science
Foundation grant \#DMS-0908379.}
\thanks{This conference note is based on the preprints~\cite{FGL12,FGL12a}.}
}

\maketitle

\begin{abstract}
It is shown that a trellis realization can be locally reduced if it is not state-trim, branch-trim,
proper, observable, and controllable.
These conditions are not sufficient for local irreducibility.
Making use of notions that amount to ``almost unobservability/uncontrollability'', a necessary and sufficient
criterion of local irreducibility for tail-biting trellises is presented.
\end{abstract}


\IEEEpeerreviewmaketitle

\section{Introduction}\label{S-Intro}
This paper deals with tail-biting trellis realizations for linear block codes.
They form the simplest type of realizations on graphs with cycles.
The latter have gained a lot of interest because they are often simpler
than realizations on cycle-free graphs and thus may give rise to more powerful iterative
decoding algorithms (e.g., LDPC codes).
Unfortunately, minimal realization theory turns out to be much more challenging for graphs
with cycles than for graphs on a finite conventional time axis, see \cite{Va98} and the references therein,
and even on general cycle-free graphs, see~\cite{FGL12}.
The discussion in~\cite{KoVa03} nicely illustrates the difficulties for the special case of
tail-biting trellis realizations.

In this note we will study tail-biting trellis realizations with the goal
to reduce their complexity by a local reduction process.
We first present necessary criteria for local irreducibility of such realizations.
They amount to well-known  basic trellis properties.
Secondly, we give a sufficient criterion.
This criterion is less obvious, and our approach will reveal how the normal graph dualization as derived in~\cite{Fo01}
helps to uncover defects of realizations that are otherwise not immediately evident.
In fact, it turns out to be beneficial to investigate a trellis and its dual simultaneously.

Finally, we will relate our results to the approach taken by Koetter and Vardy~\cite{KoVa03},
who narrowed down the class of potentially minimal tail-biting trellises by making use of product constructions.

We wish to mention that the results presented in Sections~\ref{S-Duality} and~\ref{S-NecCrit} generalize
straightforwardly to general normal graphs~\cite{FGL12,FGL12a}, but it is open whether Section~\ref{S-Suff}
can be generalized as well.

\medskip
We fix the following notations pertaining to tail-biting trellis realizations.

A linear block code $\CC$ over a finite field $\F$ is a subspace of a \emph{symbol sequence space} $\A = \Pi_{i=0}^{m-1} \A_i$, where each
\emph{symbol alphabet} $\A_i$ is a finite-dimensional vector space over $\F$.

A \emph{linear tail-biting trellis realization} (for short: \emph{trellis}) is a collection of
\emph{state spaces} $\SSS_i$ and local \emph{constraint codes} $\CC_i$
such that all spaces are finite-dimensional over~$\F$ and
$\CC_i\subseteq\SSS_i\times\A_i\times\SSS_{i+1}$ for all $i\in\Z_m:=\{0,\ldots,m-1\}$ (with index computation modulo~$m$).
The \emph{state sequence space} is defined as $\SSS=\prod_{i=0}^{m-1} \SSS_i$, and the elements of~$\CC_i$ are called
\emph{branches}.
The \emph{behavior} of the realization is the set $\Bf$ of all pairs  $(\ab, \sb) \in \A \times \SSS$ such that all constraints are
satisfied; i.e., $(s_i,a_i,s_{i+1}) \in \CC_i$ for all~$i$.
The code $\CC$ \emph{generated by the realization} is the set of all $\ab \in \A$ that appear in some
$(\ab, \sb) \in \Bf$.
Note that \emph{conventional trellis realizations} may simply be regarded as the special case where~$\SSS_0$ is trivial.

A trellis realization of the code~$\CC$ is \emph{observable} if for each $\ab \in \CC$ there is precisely
one pair $(\ab, \sb) \in \Bf$.
The realization is \emph{state-trim} (resp.\ \emph{branch-trim}) if each state (resp.\ branch) appears on a valid trajectory.
We say that a constraint code~$\CC_i$ is \emph{trim} if both the projections of~$\CC_i$ onto~$\SSS_i$ and onto~$\SSS_{i+1}$
are surjective.
A constraint code~$\CC_i$ is \emph{proper} if there are no nonzero branches of the form~$(s_i,0,0)$ or
$(0,0,s_{i+1})$ in~$\CC_i$.

Finally, we introduce the main concept of this note.
\begin{defi}\label{D-LocRed}
A \emph{local reduction} of a trellis is a replacement of one state space, say~$\SSS_i$, and the incident constraint
codes~$\CC_{i-1}$ and~$\CC_i$ by spaces of at most the same dimension while all other spaces remain unchanged such
that the resulting trellis represents the same code.
We call a local reduction \emph{strict} if the state space dimension is strictly reduced.
A realization is called \emph{locally irreducible} if each local reduction is isomorphic to the primal realization.
\end{defi}

Obviously, if no dimension is reduced then the terminology ``local reduction'' is not quite justified,
and ``local alteration'' may be more appropriate.
But we will see later (Theorem~\ref{T-S2O}) that a local reduction is always
either strict or results in a trellis that allows a subsequent strict local reduction so that ultimately we
achieve a net reduction of the trellis complexity.

The main tools for locally reducing a trellis are the mutually dual processes of trimming and merging.
They will be discussed in Section~\ref{S-NecCrit}.
It will be shown that they allow us to reduce a
trellis and its dual simultaneously. 

\section{Duality}\label{S-Duality}
We briefly recall the basic duality notions.

If $V$ is a finite-dimensional vector space over~$\F$, then its \emph{dual space}  $\hat{V}$ is defined to be the
linear algebra dual of~$V$.
Thus we have an inner product on~$V\times\hat{V}$.
For a subspace~$W\subseteq V$ the \emph{orthogonal space} $W^{\perp}\subseteq\hat{V}$ is defined as the space of all vectors
in~$\hat{V}$ that are orthogonal to all vectors in~$W$.
For further properties we refer to~\cite{Fo01,FGL12}.

Following the normal graph dualization introduced in~\cite{Fo01}, we define the
dual of a trellis~$T$ as the trellis with the same index set in which the symbol alphabets~$\A_i,\,\SSS_i$ are
replaced by their dual spaces $\hat{\A}_i,\,\hat{\SSS}_i$, the constraint codes~$\CC_i$ 
are replaced by their orthogonal spaces $\CC_i^{\perp}\subseteq\hat{\SSS}_i\times\hat{\A}_i\times\hat{\SSS}_{i+1}$, and the
sign of each dual state variable is inverted in one of the two constraints in which it is involved.
The \emph{Normal Graph Duality Theorem}~\cite{Fo01} states that
if~$T$ realizes a linear code~$\CC$, then its dual, denoted by~$T^{\circ}$, realizes the orthogonal code~$\CC^\perp$.

A main tool in dualization is projection/cross-section duality, for which we refer to~\cite[Thm.~7.2]{Fo01} or
\cite[Sec.~2.5]{FGL12}.
As a first application one obtains the following~\cite{GLW11a,FGL12}.

\begin{theo}\label{T-loctrimproper}
The constraint code~$\CC_i$ of a trellis~$T$ is trim if and only if the dual
constraint code~$\CC_i^\perp$ of~$T^{\circ}$ is proper.
\end{theo}

We call a realization~$T$ \emph{controllable} if the constraints ${\CC}_i^{\perp},\,i\in\Z_m$, are
linearly independent (in~$\hat{\A}\times\hat{\SSS}$).
A detailed discussion and justification of this terminology can be found in \cite{FGL12},
where also the following has been proved.

\begin{theo}\label{T-ObsContr}
A trellis~$T$ is controllable if and only if the dual trellis $T^{\circ}$ is observable.
Moreover,~$T$ is controllable if and only if $\dim \Bf = \sum_i \dim \CC_i - \dim \SSS$.
\end{theo}

\begin{exa}  \label{E-MergDual}
This example appeared first in~\cite{GLW11,GLW11a}.
Figure~1(a) shows the product trellis with symbol spaces $\A_i=\F_2$ for $i\in\Z_4$, obtained from the generators $0\underline{111}0,\,\underline{1}00\underline{10},\,\underline{01}\,\underline{101}$ with the indicated circular spans.
The dashed (resp.\ solid) lines denote symbol variable~$0$ (resp.~$1$).
The dual trellis is in Figure~1(b).
Note that we chose~$\hat{\SSS}_i=\SSS_i$ and the standard inner product for all state spaces and that no sign inverter is needed.
Both trellises are state-trim, proper, observable and controllable.
But the trellis in~1(b) is not branch-trim.
We will see in the next section that both trellises are locally reducible.
\begin{center}
    \includegraphics[height=2.2cm]{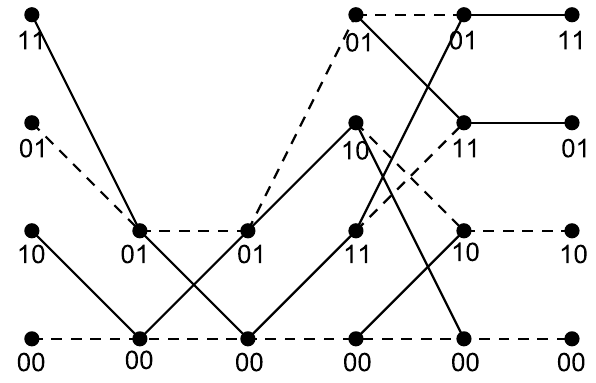}\quad \includegraphics[height=2.2cm]{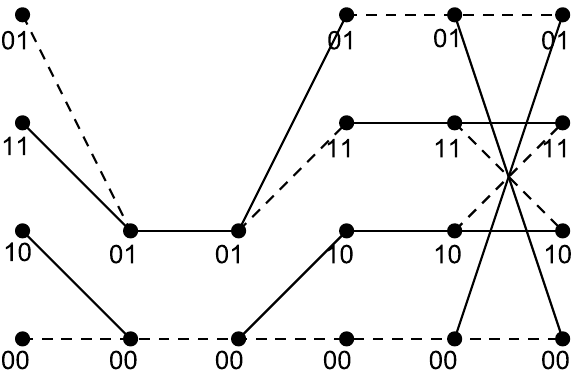}
    \\
    Figure 1(a) \hspace*{1.5cm} Figure 1(b)
\end{center}
\end{exa}

\section{Necessary Criteria for Local Irreducibility}\label{S-NecCrit}
In this section we turn to local reductions of a given trellis.
In essence, all constructions will amount to trimming (i.e., restricting) or merging (i.e., ``quotienting'') a particular state space.
Thus, let us discuss these constructions in detail.

Let~$T$ and~$T'$ be two trellis realizations of the code~$\CC$ with index set~$\{0,\ldots,m-1\}$.
Denote the state spaces and constraint codes by~$\SSS_i,\,\CC_i$ and~$\SSS_i',\,\CC_i'$, respectively.
We say that~$T'$ is \emph{obtained from~$T$ by trimming the state space~$\SSS_i$}, if
$\SSS_i'$ is a subspace of~$\SSS_i$ and
$\CC_{i-1}'=\{(s_{i-1},a_{i-1},s_i)\in\CC_{i-1}\mid s_i\in\SSS_i'\}$ and
$\CC_{i}'=\{(s_i,a_i,s_{i+1})\in\CC_{i}\mid s_i\in\SSS_i'\}$, while all other state spaces and
constraint codes of~$T$ and~$T'$ coincide.

The dual of state-trimming is an instance of what is known as merging in the literature of trellis realizations.
In our case, merging has to be done by taking a suitable quotient space in order to keep the linearity of the resulting trellis.
Precisely, we say the realization~$T'$ \emph{is obtained from~$T$ by merging at time~$i$} if $\SSS_i'=\SSS_i/\cY$ for
some subspace $\cY\subseteq\SSS_i$ and if
$\CC_{i-1}'=\{(s_{i-1},a_{i-1},s_i+\cY)\mid (s_{i-1},a_{i-1},s_i)\in\CC_{i-1}\}$ and similarly
$\CC_i'=\{(s_i+\cY,a_i,s_{i+1})\mid (s_i,a_i,s_{i+1})\in\CC_i\}$.
Again, all other state spaces and constraint codes of~$T'$ and~$T$ coincide.
It follows from projection/cross-section duality that~$T'$ is obtained from~$T$ by trimming~$\SSS_i$ to
the subspace~$\cY$ if and only if~$(T')^{\circ}$ is obtained from~$T^{\circ}$ by merging~$\hat{\SSS}_i$ to
$\hat{\SSS}_i/\cY^{\perp}$.

In general, the trimmed/merged realization generates a different code than the original realization.
We will, of course, be interested in the case where the code generated does not change after trimming/merging.
In this case, trimming and merging obviously form simultaneous local reductions of the trellis and its dual in the sense of Definition~\ref{D-LocRed}.

Now we can present a first list of necessary conditions for local irreducibility.

\begin{theo}\label{T-LocIrrNec}
Any locally irreducible realization is state-trim, proper, observable, and controllable.
More precisely, if a realization~$T$ lacks any of these properties, then there exists a
strict local reduction of~$T$, and the dual process is a strict local reduction of~$T^{\circ}$.
\end{theo}
\begin{IEEEproof}
If~$T$ is not state-trim or not proper, then we can trim either~$T$ or~$T^{\circ}$
(see Theorem~\ref{T-loctrimproper}).
The dual process is a merging, and we obtain strict local reductions for~$T$ and~$T^{\circ}$.
\\
Next, suppose~$T$ is unobservable.
Precisely, let $(\zerob, \sb)$ be a non-trivial unobservable trajectory and let~$s_i\in\SSS_i$ be a nonzero state
on this trajectory.
Let $\cT_i$ be a subspace of~$\SSS_i$ such that $\cT_i\oplus\inner{s_i}=\SSS_i$.
It is not hard to see (\cite{FGL12}) that the realization obtained by trimming~$\SSS_i$ to~$\cT_i$ still generates the same code.
Thus we have obtained a strict local reduction.
The dual process is a suitable state merging of~$T^{\circ}$, and this again is a strict local reduction.
This also shows that uncontrollable realizations are locally reducible.
\end{IEEEproof}

\medskip
Note that the trellises in Figure~1 have all the necessary conditions of the previous theorem, and thus this
simple test does not imply their local reducibility.
In fact, we need to make use of the fact that the trellis in Figure~1(b) is not branch-trim.

The following notion will be crucial for our further investigation.
It addresses non-branch-trimness in a way that is naturally dualizable.
This will allow us to uncover the shortcomings of the trellis in Figure~1(a).

\begin{defi}\label{D-S1O}
An observable and controllable trellis is called \emph{span-one-un\-con\-trol\-lable} (resp.\ \emph{span-one-unobservable}) if we can
replace one constraint code by a proper subspace (resp.\ super\-space) without changing the code realized by the trellis.
\end{defi}

Evidently, an observable and controllable trellis is span-one-uncontrollable if and only if it is not branch-trim.
Moreover, a trellis~$T$ is span-one-uncontrollable if and only if its dual~$T^{\circ}$ is span-one-unobservable.
This follows immediately from the equivalence $\tilde{\CC}_i\subset\CC_i\Leftrightarrow\tilde{\CC}_i^{\perp}\supset\CC_i^{\perp}$,
which also shows that replacing a constraint code by a proper subspace or superspace are mutually dual processes.
Furthermore, if~$T$ is span-one-uncontrollable, then a branch-trimming (i.e., replacing a constraint code by a proper
subcode) results in an uncontrollable trellis.
This is a direct consequence of Theorem~\ref{T-ObsContr}.
Dually, a span-one-unobservable trellis will become unobservable after branch-addition (replacing a constraint code by a proper supercode).
This explains the terminology.

\medskip
Now we can formulate another necessary condition for local irreducibility.

\begin{theo}\label{T-LocIrrS1O}
Let~$T$ be span-one-unobservable. Then~$T$ allows a strict local reduction.
The dual process is a strict local reduction of~$T^{\circ}$, and thus
span-one-uncontrollable trellises are locally reducible as well.
\end{theo}

\begin{IEEEproof}
Let~$T$ be span-one-unobservable (but observable).
Then~$T^{\circ}$ is span-one-uncontrollable and we may replace a constraint code, say~$\CC_i^{\perp}$, by a strict
subspace, say~$\tilde{\CC}_i^{\perp}$, without changing the code generated by the realization.
Without loss of generality we may assume $\dim\tilde{\CC}_i^{\perp}=\dim\CC_i^{\perp}-1$.
Theorem~\ref{T-ObsContr} shows that the resulting realization, denoted by~$\tilde{T}^{\circ}$, is uncontrollable.
The dual process is the expansion of the constraint code~$\CC_i$ to the superspace $\tilde{\CC}_i$ where
$\dim\tilde{\CC}_i=\dim\CC_i+1$, and by Theorem~\ref{T-ObsContr} we obtain an unobservable
realization~$\tilde{T}$.
Thus by Theorem~\ref{T-LocIrrNec} the trellis~$\tilde{T}$ can be trimmed.
We have to do this carefully in order to decrease the dimension of the enlarged constraint code~$\tilde{\CC}_i$.
Since~$T$ is observable, but $\tilde{T}$ is not, there exists a nonzero branch
$(s_i,0,s_{i+1})\in\tilde{\CC}_i\backslash\CC_i$.
Without loss of generality, let $s_i\neq0$ and thus we may trim the state space~$\SSS_i$
as described in the proof of Theorem~\ref{T-LocIrrNec}.
The resulting realization, denoted by~$\hat{T}$, has a strictly smaller state space and constraint code
at time~$i$ than~$\tilde{T}$.
Hence the final constraint code dimensions of~$\hat{T}$ are not larger than those of~$T$, and
we have performed a strict local reduction of~$T$.
The dual process is a strict local reduction of~$T^{\circ}$ because it
is a branch-trimming followed by a state-merging.
\end{IEEEproof}

\medskip
Now we can reduce the trellises in Figure~1.

\medskip
\noindent{\bf Example~\ref{E-MergDual}} (cont.)
The trellis in Figure~1(b) is span-one-uncontrollable.
Indeed, the trellis is not branch-trim, and deleting the four diagonal branches of the last constraint code
does not change the code generated by that trellis.
Thus, the trellis and its dual in Figure~1(a) are locally reducible.
Deleting those four diagonal branches renders the uncontrollable trellis in Figure 2(b).
The dual process consist of expanding the last constraint code in the trellis in Figure~1(a) by one dimension, namely by adding
the branch $(01|0|01)$ to the generators of this constraint code.
This results in the unobservable trellis in Figure 2(a), which then is the dual of that in Figure 2(b).
\begin{center}
    \includegraphics[height=2.2cm]{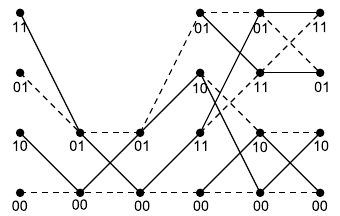}\quad \includegraphics[height=2.2cm]{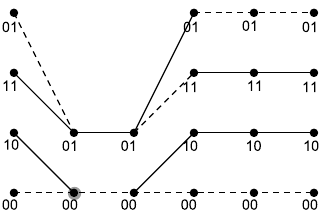}
    \\
    Figure 2(a) \hspace*{1.5cm} Figure 2(b)
\end{center}
Now we can trim the first trellis at state time~$4$ to the subspace $\{00,11\}$ and merge the dual state space to
$\F_2^2/\{00,11\}$.
This results in the mutually dual trellises shown in Figure~3.
These trellises still generate the original code~$\CC=\inner{01110,\,10010,\,01101}$ and its dual~$\CC^{\perp}$, respectively.
Notice that they form strict local reductions of the trellises in Figure~1 in the sense of Definition~\ref{D-LocRed}.
\begin{center}
    \includegraphics[height=2.2cm]{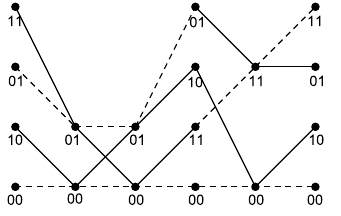}\quad \includegraphics[height=2.2cm]{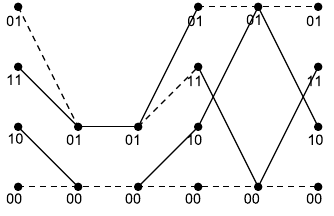}
    \\
    Figure 3(a) \hspace*{1.5cm} Figure 3(b)
\end{center}
Is is worth emphasizing that the trellis in Figure~1(a) is locally reducible, even though it is
state-trim, proper, branch-trim, observable, controllable, and non-mergeable (i.e., no state space can be merged without
changing the code generated by the trellis).
We first had to perform an auxiliary branch-addition before a state-trimming resulted in a net reduction of the trellis complexity.

\medskip
The necessary conditions for local irreducibility we have presented so far are not sufficient.
For instance, the trellises in Figure~3 are state-trim, branch-trim, proper, span-one-observable, and span-one-controllable.
Yet they are still locally reducible, as we will see in the next section.

\section{Sufficient Condition of Local Irreducibility}\label{S-Suff}
In this section we will present a characterization of local irreducibility of tail-biting trellises.
For ease of notation, we will restrict ourselves to trellis realizations with symbol spaces $\A_i=\F$ for all~$i$.
The results can straightforwardly be generalized to arbitrary trellises.
Following the usual notation, all trellises will have length~$n$ and thus generate codes in~$\F^n$.

Let us return to span-one-unobservable trellises. We have the following alternative description.
\begin{prop}\label{P-S1O}
An observable and controllable trellis is span-one-un\-ob\-ser\-vab\-le if and only if there exists a nontrivial path of length~$n-1$
representing the zero word (of length~$n-1$).
\end{prop}
\begin{IEEEproof}
It is clear from the paragraph after Definition~\ref{D-S1O} that a span-one-unobservable trellis must contain such a path.
Conversely, let~$T$ be observable and controllable with behavior~$\Bf$, and let~$T$ contain a nontrivial
path of length~$n-1$ representing the zero word.
Suppose the path passes through the states $s_1,s_2,\ldots,s_0$.
Replace the constraint code~$\CC_0$ by $\CC_0+\inner{(s_0|0|s_1)}$.
Then the new realization contains~$\Bf$ and the new unobservable trajectory
$(\zerob,\sb)$, where $\sb=(s_0,\ldots,s_{n-1})$.
But since the constraint code increases by just one dimension, the behavior can also increase by
at most one dimension.
Thus, the behavior is given by $\Bf+(\zerob,\sb)$, and the new trellis still generates the same code.
This shows that~$T$ is span-one-unobservable.
\end{IEEEproof}

\medskip
All of this suggests to extend the notion of span-one-observability to longer spans.
However, for our purpose of local reductions it will suffice to consider span-two-observability/controllability.
\begin{defi}\label{D-S2O}
A trellis realization of length~$n$ is called \emph{span-two-observable} if it does not contain a nontrivial path of
length~$n-2$ representing the zero word.
A trellis is called \emph{span-two-controllable} if its dual is span-two-observable.
\end{defi}

Evidently span-two-observability implies span-one-ob\-ser\-va\-bi\-li\-ty implies observability.
Span-two-controllability can be characterized as follows.
\begin{prop}\label{P-S2C}
A trellis is span-two-controllable if and only if, for each $i\in\Z_m$, each pair of states in~$\SSS_i$ and~$\SSS_{i-2}$ are connected by a path of length~$n-2$.
\end{prop}
\begin{IEEEproof}
Without loss of generality let $i=2$.
Let~$\cP$ be the set of all paths in the trellis~$T$ starting at time~$2$ and ending at time~$0$.
Then the projection $\cP|_{\SSS_2\times\SSS_0}$ is the set of all state pairs that are connected by such a path.
With the aid of projection/cross-section duality one obtains that
$(\cP|_{\SSS_2\times\SSS_0})^{\perp}$ is the space of all pairs in $\hat{\SSS}_2\times\hat{\SSS}_0$ that are connected by a
path in~$T^{\circ}$ that represents the zero word.
Thus, $T^{\circ}$ is span-two-observable if and only if this space is trivial, which in turn is equivalent to
$\cP|_{\SSS_2\times\SSS_0}=\SSS_2\times\SSS_0$.
This proves the desired result.
\end{IEEEproof}

\medskip
Now we can formulate our main result characterizing local irreducibility.
\begin{theo}\label{T-S2O}
Let~$\CC\subseteq\F^n$ be a code such that both~$\CC$ and~$\CC^{\perp}$ do not contain any codewords that are
supported by an interval of length at most~$2$.
Let~$T$ be a trellis realization of~$\CC$.
Then the following are equivalent.
\begin{romanlist}
\item $T$ is locally irreducible.
\item $T$ is state-trim, branch-trim, proper, span-two-observable and span-two-controllable.
\item $T^{\circ}$ is locally irreducible.
\item $T^{\circ}$ is state-trim, branch-trim, proper, span-two-observable and span-two-controllable.
\end{romanlist}
Furthermore, if~$T$ is locally reducible, then it allows either a strict local reduction or a local reduction which
gives rise to a subsequent strict local reduction.
In either case, the dual process is a reduction of the same type for~$T^{\circ}$.
\end{theo}

\begin{IEEEproof}
We have seen already in Theorems~\ref{T-LocIrrNec} and~\ref{T-LocIrrS1O} that if a trellis is not state-trim, branch-trim, proper,
span-one-observable/controllable, then it and its dual are locally reducible.
If a trellis has all these properties, but is not span-two-observable, then a local reduction is illustrated in Example~\ref{E-KVSCP} below.
The detailed proof can be found in~\cite{FGL12a}.

Let us sketch the proof for (ii)~$\Rightarrow$~(i). The remaining statements follow easily.
Thus, let~$T$ be a trellis as in~(ii).
Suppose~$\tilde{T}$ is a local reduction of~$T$ with state space~$\tilde{\SSS}_1$ and constraint codes~$\tilde{\CC}_0$ and~$\tilde{\CC}_1$,
while all other spaces of~$\tilde{T}$ and~$T$ coincide.
We have to show that~$T$ and~$\tilde{T}$ are isomorphic, that is, they coincide up to state space isomorphisms.
Without loss of generality we may assume that~$\tilde{T}$ is trim and proper at state time~$1$, and thus at all times.
One can easily see that~$\tilde{T}$ and~$\tilde{T}^{\circ}$ are observable.
Let $\dim\tilde{\SSS}_1=\dim\SSS_1-m$ and $\dim\tilde{\CC}_i=\dim\CC_i-\ell_i$ for $i=0,1$, where $m,\,\ell_o,\ell_1\geq 0$.
With the aid of Theorem~\ref{T-ObsContr} one derives straightforwardly $\ell_0+\ell_1=m$.

We show next that $\ell_o=\ell_1=m=0$.
To this end, let $(s_0,a_0,s_1)\in\CC_0$.
This branch belongs to a valid trajectory in~$T$, say $(\ab,\sb)$.
By observability, there is a unique valid trajectory $(\ab,\tilde{\sb})$ in~$\tilde{T}$,
and thus we obtain a branch $(s_0,a_0,\tilde{s}_1)\in\tilde{\CC}_1$.
Using properness of~$T$ and~$\tilde{T}$, this gives rise to
a well-defined, injective linear map $\varphi_0:\CC_0\rightarrow\tilde{\CC}_0$, $(s_0,a_0,s_1)\mapsto (s_0,a_0,s'_1)$.
As a consequence, $\dim\tilde{\CC}_0\geq\dim\CC_0$.
But then we have equality and~$\varphi_0$ is an isomorphism.
In the same way $\dim\tilde{\CC}_1=\dim\CC_1$ and there is an isomorphism
$\varphi_1:\CC_1\rightarrow\tilde{\CC}_1,\ (s_1,a_1,s_2)\mapsto (s''_1,a_1,s_2)$.
Now $\ell_0+\ell_1=0$ implies $m=0$, thus $\dim\tilde{\SSS}_1=\dim\SSS_1$, and the given local reduction is not strict.

Finally, by Proposition~\ref{P-S2C}, for each $s_2\in\SSS_2$ and $s_0\in\SSS_0$ there exists a path in~$T$
starting in~$s_2$ and ending in~$s_0$.
As a consequence, each path $(s_0,a_0,s_1,a_1,s_2)$ of length~$2$ in~$T$ is part of a valid trajectory in~$T$.
This in turn yields that for each such path $(s_0,a_0,s_1,a_1,s_2)$ there exists some~$\tilde{s}_1\in\tilde{\SSS}_1$ such that
$\varphi_0(s_0,a_0,s_1)=(s_0,a_0,\tilde{s}_1)$ and $\varphi_1(s_1,a_1,s_2)=(\tilde{s}_1,a_1,s_2)$.
This way we obtain a well-defined isomorphism~$\psi:\SSS_1\rightarrow\tilde{\SSS}_1$
such that $\tilde{\CC}_0=\{(s_0,a_0,\psi(s_1))\mid (s_0,a_0,s_1)\in\CC_0\}$ and similarly for~$\tilde{\CC}_1$.
But this simply means that~$\tilde{T}$ and~$T$ are isomorphic trellises.
\end{IEEEproof}

\begin{exa}\label{E-KVSCP}
Consider the trellises given in Figure~4 below.
The trellis in Figure~4(a) is the product trellis of the code~$\CC$ generated by
$\underline{11001}0,\,\underline{001}\,\underline{110},\,\underline{1}000\underline{11}$ with the
indicated spans.
(The code~$\CC^{\perp}$ does not satisfy the requirements of Theorem~\ref{T-S2O}.
But the example still illustrates the main arguments of the general proof; this is due to the fact
that the reduction is based on span-two-unobservability and not span-two-uncontrollability).
The trellis is state-trim, branch-trim, proper, observable, and controllable.
It is also span-one-observable and span-one-controllable, but not span-two-observable: it contains a
path starting at the state $11\in\SSS_4$, ending at $10\in\SSS_2$, and representing the zero word.
This will allow us to perform a local reduction by changing the state space~$\SSS_3$ and its incident
constraint codes~$\CC_2,\,\CC_3$.
In a first step we expand~$\SSS_3$ by introducing an additional state, say
$\tilde{\SSS}_3:=\SSS_3+\inner{\tilde{s}}$, and expand~$\CC_2$ and~$\CC_3$ to
$\tilde{\CC}_2=\CC_2+\inner{(10|0|\tilde{s})}$ and $\tilde{\CC}_3=\CC_3+\inner{(\tilde{s}|0|11)}$.
It is not hard to see that the resulting trellis, shown in Figure~4(b), still represents the code~$\CC$.
By construction, it is unobservable and thus can be trimmed as in the proof of Theorem~\ref{T-LocIrrNec}.
If we trim the state space~$\tilde{\SSS}_3$ to the subspace~$\inner{1+\tilde{s}}$, we obtain the
trellis in Figure~4(c).
This trellis forms a local reduction of the original trellis where no state space and constraint code
dimension has changed.
But the new trellis is obviously not trim, and thus we may trim it at state time~$4$, resulting in
the trellis shown in Figure~4(d).
This illustrates the main idea of the proof of (i)~$\Rightarrow$~(ii) in Theorem~\ref{T-S2O}: one can show
that for a span-two-unobservable trellis there always exists an extension of one state space followed by a trimming
that results in a non-trim trellis, which then can be further reduced.

It is easy to see that dualizing the preceding steps leads to a two-step local reduction of the dual trellis.

The trellis in Figure~4(d) is state-trim, branch-trim, proper, span-two-observable,
span-one-controllable, but not span-two-controllable.
Thus it can be further locally reduced.
\begin{center}
    \includegraphics[height=2cm]{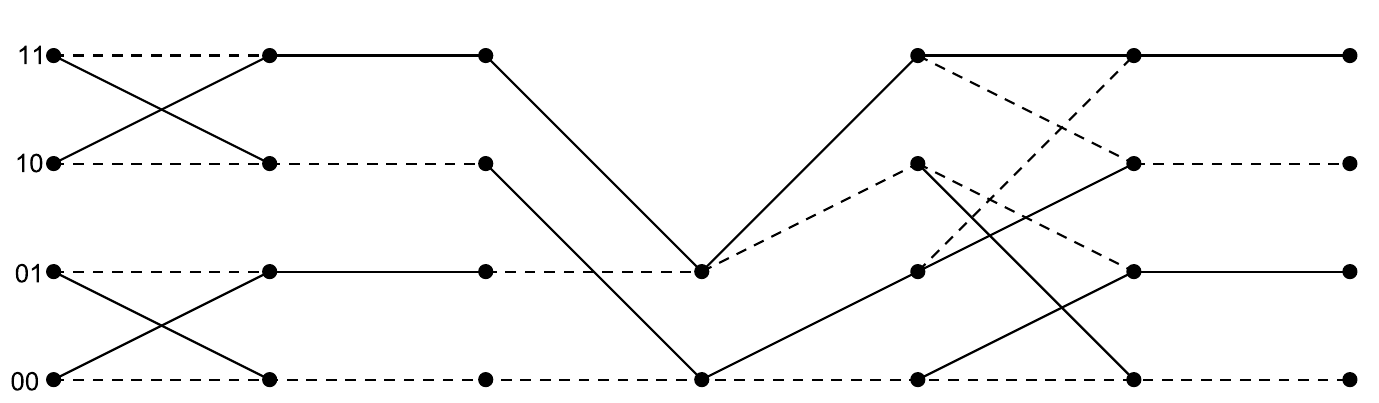}
    \\
    Figure 4(a)
    \\
    \includegraphics[height=2cm]{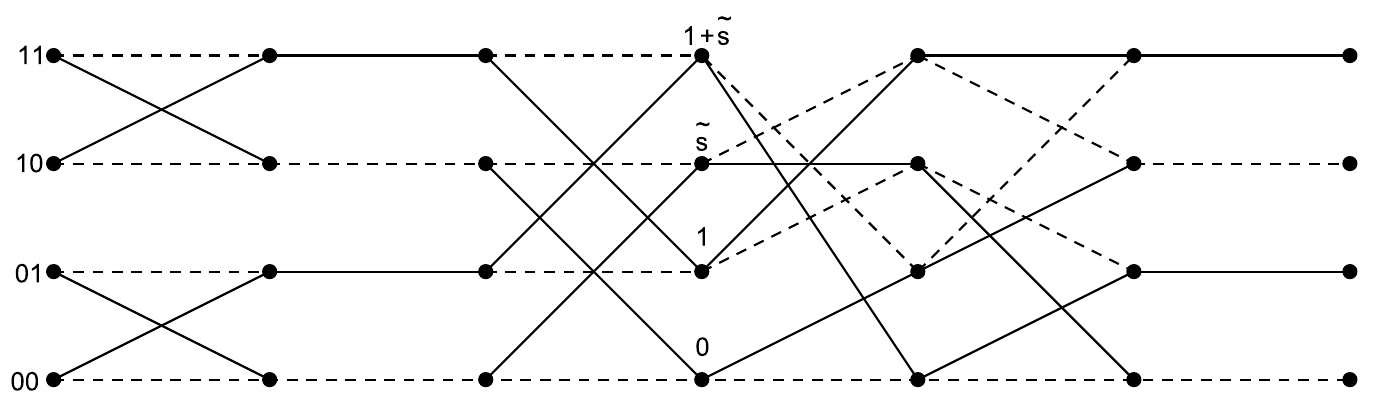}
    \\
    Figure 4(b)
    \\
    \includegraphics[height=2cm]{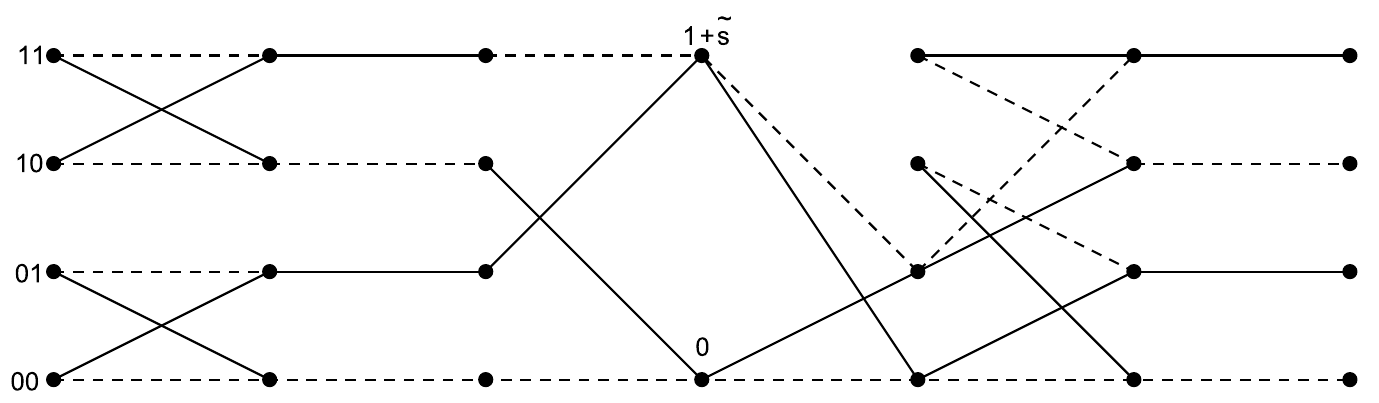}
    \\
    Figure 4(c)
    \\
    \includegraphics[height=2cm]{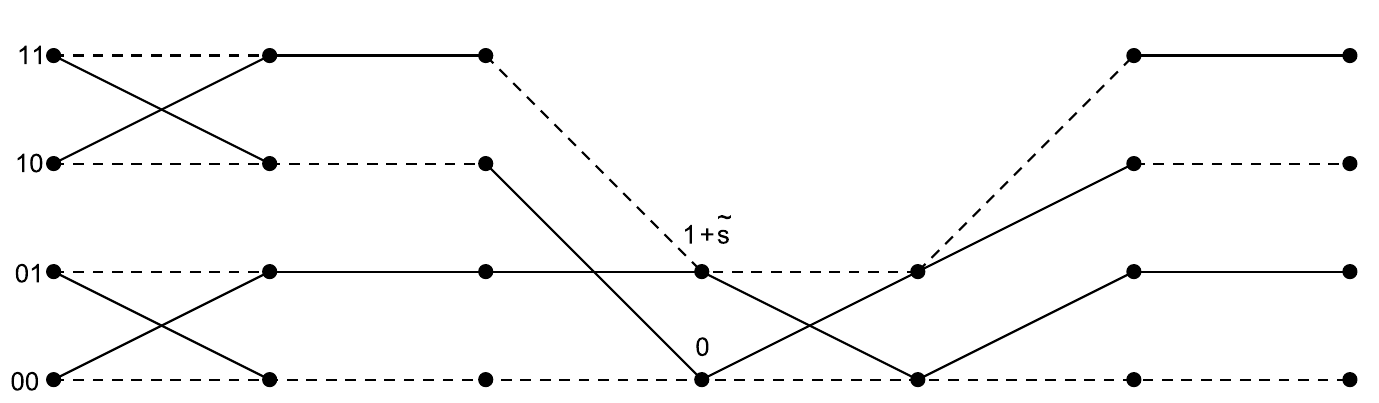}
    \\
    Figure 4(d)
\end{center}

In the same manner the trellis in Figure~3(a) can be locally reduced.
\end{exa}

\section{Summary for tail-biting trellises}
We have given a complete description of tail-biting trellises that are locally irreducible in the sense of
Definition~\ref{D-LocRed}.
Our methods were solely based on intrinsic trellis properties, and no specific representation like the product construction was used.
Furthermore, we provided constructive procedures of how to reduce a given trellis (and its dual).

Let us compare our results with previous work on tail-biting trellises with potentially low complexity.
To this end, we restrict ourselves again to trellises with symbol spaces~$\A_i=\F$ for all~$i$.
Koetter and Vardy \cite{KoVa03} showed that the search for possibly minimal linear tail-biting trellis realizations
could be narrowed to a certain canonical class, which we call \emph{KV-trellises}
(here minimal refers to $\Theta$-minimal in the sense of~\cite{KoVa03}).
A KV-trellis is a product realization with $\dim \CC$ linearly independent generators, each of which has the
shortest circular span of any codeword whose span starts (or stops) at the same position \cite{GLW11,GLW11a}.
The construction may be regarded as the tail-biting version of the ``shortest basis" approach~\cite{Fo11a}.
KV-trellises are state-trim, branch-trim, proper, observable and controllable~\cite{KoVa03}.

Now we can give an overview of various trellis classes.
Let us write S1O, S2O, S1C, S2C for span-one-obser\-va\-bi\-li\-ty, span-two-observability, and the dual controllability notions.
We also denote ``state-trim, branch-trim and proper'' by T/P, and ``observable and controllable" by O/C.

With the machinery developed in~\cite{GLW11a}, one can show that if a code does not contain
any span-two codewords, then its KV-trellises are S2O.
Thus, if both the code and its dual do not contain any span-two codewords (a requirement that is not met by the examples
in this note), then their KV-trellises are S2O and S2C.
On the other hand, there exist examples of T/P and S2O/S2C trellises that are not KV.
All of this leads to the following strict containments of classes of tail-biting trellises for a given code
(assuming that neither the code nor its dual contains any span-two codewords):
\[
 \begin{array}{l}\{\textrm{minimal}\}\subsetneq\{\textrm{KV}\}\subsetneq\{\textrm{T/P, S2O/S2C}\}\subsetneq
  \{\textrm{T/P, S1O/S1C}\}\\[.5ex]
  \quad\subsetneq\{\textrm{T/P, O/C}\}.\end{array}
\]
Furthermore, the four trellis classes in the top row are invariant under taking duals, whereas the one in the bottom row
is not. The set $\{\textrm{T/P, S2O/S2C}\}$ is exactly the class of locally irreducible trellises.



\bibliographystyle{abbrv}
\bibliography{literatureAK,literatureLZ}

\begin{thebibliography}{1}

\bibitem{Fo01}
{G.~D. Forney, Jr.}
\newblock Codes on graphs: {N}ormal realizations.
\newblock {\em IEEE Trans. Inform. Theory}, IT-47:520--548, 2001.

\bibitem{Fo11a}
{G.~D. Forney, Jr.}
\newblock Minimal realizations of linear systems: {T}he ``shortest basis''
  approach.
\newblock {\em IEEE Trans. Inform. Theory}, IT-57:726--737, 2011.

\bibitem{FGL12}
{G.~D. Forney, Jr.} and H.~Gluesing-Luerssen.
\newblock Codes on graphs: {O}bservability, controllability and local
  reducibility.
\newblock Preprint, 2012.

\bibitem{FGL12a}
H.~Gluesing-Luerssen and {G.~D. Forney, Jr.}
\newblock Local reducibility of tail-biting trellises.
\newblock In preparation, 2012.

\bibitem{GLW11}
H.~Gluesing-Luerssen and E.~Weaver.
\newblock Linear tail-biting trellises: {C}haracteristic generators and the
  {BCJR}-construction.
\newblock {\em IEEE Trans. Inform. Theory}, IT-57:738--751, 2011.

\bibitem{GLW11a}
H.~Gluesing-Luerssen and E.~Weaver.
\newblock Characteristic generators and dualization for tail-biting trellises.
\newblock {\em IEEE Trans. Inform. Theory}, IT-57:7418--7430, 2011.

\bibitem{KoVa03}
R.~Koetter and A.~Vardy.
\newblock The structure of tail-biting trellises: {M}inimality and basic
  principles.
\newblock {\em IEEE Trans. Inform. Theory}, IT-49:2081--2105, 2003.

\bibitem{Va98}
A.~Vardy.
\newblock Trellis structure of codes.
\newblock In {\em Handbook of Coding Theory, Vol.~2 {\rm (V.~S.~Pless and
  W.~C.~Huffman, eds.)}}, pages 1989--2117. Elsevier, Amsterdam, 1998.

\end{thebibliography}
\end{document}